\documentclass{PoS}

\usepackage{amsmath}% for dealing with mathematics,
\usepackage{amsthm}% for dealing with theorem environments,
\usepackage{graphicx}% for dealing with figures.
\usepackage{subfigure}
\usepackage{bm}
\usepackage{booktabs}
\usepackage{dcolumn}

\newcommand{\nn}{\nonumber\\}
\graphicspath{{./figs/}}

\title{A non-perturbative effect of gluons for scalar diquark 
in the Schwinger-Dyson formalism}

\ShortTitle{A non-perturbative effect of gluons for scalar diquark 
in the Schwinger-Dyson formalism}

\author{\speaker{Shotaro Imai}, Hideo Suganuma\\
Department of Physics, Graduate School of Science,
  Kyoto University, \\
  Kitashirakawa-oiwake, Sakyo, Kyoto 606-8502, Japan\\
        E-mail: \email{imai@ruby.scphys.kyoto-u.ac.jp}}

\abstract{
The diquark has been considered to be important effective degree of freedom in hadron physics, especially for multi-quark physics and the structure of heavy hadronic states. Using the Schwinger-Dyson formalism, we investigate the non-perturbative effect of gluons for a scalar diquark with renormalization-group improved coupling in the Landau gauge. Here, we treat the scalar diquark as an effective degree of freedom with a peculiar size, while the diquark is originally a bound-state-like object of two quarks. Since the diquark has a non-zero color charge, it still strongly interacts with gluons. We evaluate the gluonic non-perturbative effect to the diquark, considering the size effect of the diquark. We investigate the mass function of the diquark in both cases with a constant bare diquark mass and twice of the running quark self-energy. It is found that the diquark, especially the small diquark, obtains a large effective mass by the gluonic dressing effect. The scalar diquark mass seems to be dynamically generated by the non-perturbative effect, although it does not have chiral symmetry explicitly.}

\FullConference{XV International Conference on Hadron Spectroscopy-Hadron 2013\\
		4-8 November 2013\\
		Nara, Japan }

\begin{document}

\section{Introduction}

Recent experiments have discovered many heavy hadronic states including exotic 
states, which cannot be understood as ordinary hadrons~\cite{Beringer2012}. 
In order to describe the structure of their states, the diquark is considered 
as an effective degree of freedom. For the diquark, the most attractive 
channel by the one-gluon exchange 
is the color and flavor anti-triplet $\bm{\bar{3}}_{c,f}$ and 
spin singlet with even parity $0^{+}$.
If the diquark correlation is developed in a hadron such as a heavy baryon 
($Qqq$), this scalar diquark channel would be favored.
The diquark is made by two quarks with gluonic interaction, and it still strongly 
interacts with gluons because of its non-zero color charge. 
The dynamics of diquark and gluons may affect the structure of a hadron.

In quark-hadron physics, the Schwinger-Dyson (SD) formalism 
is often used to evaluate the non-perturbative effect based on QCD 
\cite{Higashijima1984,Aoki1990,Yamanaka2013}. 
We apply the SD formalism to the scalar diquark with its peculiar size, 
and investigate the effective diquark mass, 
which reflects a non-perturbative dressing effect by gluons. 
The diquarks are sometimes treated as point-like object or local boson fields. 
However, the diquark must have an effective size, 
since it is a bound-state-like object inside a hadron. 
Therefore, we investigate also the size effect of the diquark for its dynamics.

%The properties of diquarks, however, such as the size and the mass are not understood.
%  Since the diquark belongs the color representation $\bm{\bar{3}}_c$, it strongly interacts with gluon and is confined in a hadron. 
% The diquark in a hadron may have effective size and affect the structure of hadrons.
%Thus, the dynamics of diquark and gluon is important to understand the properties of diquark and structure of exotic hadrons. 

% In order to evaluate the non-perturbative effect, the Schwinger-Dyson equation is often used. We apply the formalism to the scalar diquark with peculiar size.

\section{The Schwinger-Dyson Equation for the Scalar Diquark}

We consider the scalar diquark as an effective degree of freedom 
with a peculiar size  like a meson, assuming 
%it to be an extended fundamental field $\phi(x)$ like a meson in effective hadron models. 
it to be an extended fundamental field $\phi(x)$~\cite{Imai2014}. 
The scalar diquark interacts with gluons since it has non-zero color charge 
and is affected by non-perturbative gluonic effects~\cite{Iida2007}.

% , which dynamics is described by the Lagrangian:
% \begin{align}
%  \mcl{L}=[(\partial^\mu-igA^{\mu}_{a}T^{a})\phi]^\dagger[(\partial_{\mu}-igA_{\mu b}T^{b})\phi]-m_{\phi}^2\phi^\dagger\phi,
% \end{align}
% where the bare diquark mass $m_{\phi}$ and the gauge field $A^{\mu}_{a}$ with the generator $T^a$ ($a=1,\cdots,8$) have been introduced. %We note that the scalar diquark has the 4-point interaction term different from the quark (fermion).
To evaluate the non-perturbative effect, 
we take the Schwinger-Dyson (SD) formalism  
%The Schwinger-Dyson (SD) formalism is used in this study.
with the rainbow-ladder truncation with the Higashijima-Miransky 
approximation $\alpha_s((p_E-k_E)^2) \approx \alpha_s(\max(p_E^2,k_E^2))$ 
in the Landau gauge.
Here, $p_E$ denotes the Euclidean momentum.
We use a renormalization-group(RG)-improved coupling 
in the case of $N_c=3$ and $N_f=3$,
\begin{align}
 \alpha_s(p_E^2)=\frac{g^2(p_E^2)}{4\pi}=
\begin{cases}
 \frac{12\pi}{11N_c-2N_f}\frac{1}{\ln(p_E^2/\Lambda_{\rm QCD}^2)} 
\quad (p_E^2 \geq p_{\rm IR}^2)\\
 \frac{12\pi}{11N_c-2N_f}\frac{1}{\ln(p_{\rm IR}^2/\Lambda_{\rm QCD}^2)} 
 \quad (p_E^2 \leq p_{\rm IR}^2)
\end{cases},\label{eq:coupling}
\end{align}
with an infrared regularization of a simple cut at $p_{\rm IR}\simeq 640$ MeV which leads to $\ln(p_{\rm IR}^2/\Lambda_{\rm QCD}^2)=1/2$, and the QCD scale parameter $\Lambda_{\rm QCD}=500$ MeV~\cite{Aoki1990,Yamanaka2013}.
%
%The infrared regularization has been introduced to avoid the divergent pole at $p=\Lambda_{QCD}$.
%The scale parameter $\Lambda_{QCD}$ is chosen to reproduce the chiral properties for quark in the SD formalism with the Higashijima-Miransky approximaion in the Landau gauge, while the ordinary QCD scale parameter is around $\Lambda_{QCD}\sim 200-300$ MeV.
%It is notable that we can use the same form of the running coupling for quark-gluon coupling even for the scalar-gluon. The origin of the running coupling comes from the renormalization group analysis. In the heavy mass limit of interacting particles, the interaction depends on only color. The scalar diquark can be identified as anti-quark in the leading order. %Hence, we can use same running coupling even for scalar-gluon case.
%As for the running coupling, the fundamental behavior should be same because which comes from the renormalization group argument for the quark-gluon vertex.
%
%Since our diquark is confined and located in a hadron, the diquark may have an effective size with smaller than the hadron. 
%
To include the size effect of diquark, 
we introduce a simple form factor in the 4D Euclidean space as
\begin{align}
 f_{\Lambda}(p_E^2)=\left(\frac{\Lambda^2}{p_E^2+\Lambda^2}\right)^2,
\end{align}
where the momentum cutoff $\Lambda$ corresponds to 
the inverse of the diquark size $R$, and we here set $ R \equiv \Lambda^{-1}$.
% It is known that the radiative correction of scalar particle is diverge even in perturbation theory ({\itshape ref. Higgs?}). This form factor has a role of the convergence factor at the same time. 
The size effect of the diquark can be included 
in the vertex as $\alpha_s(p^2)\to\alpha_s(p^2)f_{\Lambda}(p^2)$.

\begin{figure}[htbp]
\centering
   \includegraphics[width=1\textwidth]{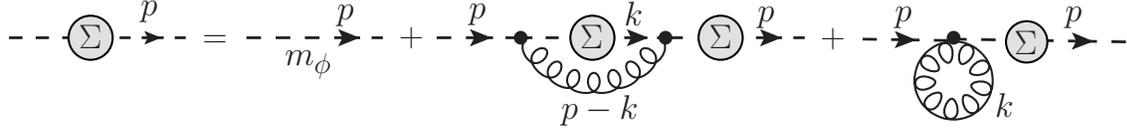}
 \caption{The Schwinger-Dyson equation for the scalar diquark. The shaded blob is the self-energy $\Sigma(p^2)$, the dashed line denotes the scalar diquark propagator and the curly line the gluon propagator. The last diagram in RHS is peculiar term of scalar theory.}\label{fig:sd_d} % The last term (4-point interaction) is the peculiar term of scalar theory, which does not appear in the single quark case.
\end{figure}

The SD equation for the scalar diquark with the bare diquark mass $m_{\phi}$ is diagrammatically expressed 
as Fig.~\ref{fig:sd_d} and is written by
% The SD equation for the scalar diquark is diagrammatically expressed 
% as Fig.~\ref{fig:sd_d} and is written by
\begin{align}
 \Sigma^2(p_E^2)% =&m_{\phi}^2+\frac{3C_{2}(\bm{3})}{2\pi^3}\int_{0}^{\Lambda_{UV}} d^4k_E \frac{\alpha_s(k_E^2)f_{\Lambda}(k_E^2)}{k_E^2}\nn
% &-\frac{C_{2}(\bm{3})}{4\pi^3}\int_{0}^{\Lambda_{UV}} d^4k_E\frac{\alpha_s((p_E-k_E)^2)f_{\Lambda}((p_E-k_E)^2)}{k_E^2+\Sigma^2(k_E^2)}\frac{4p_E^2k_E^2-4(p_E\cdot k_E)^2}{(p_E-k_E)^4}\nn
=&m_{\phi}^2+\frac{8}{\pi}\int_{0}^{\infty} dk_E k_E \alpha_s(k_E^2)f_{\Lambda}(k_E^2)\nn
&-\frac{2\alpha_{s}(p_E^2)f_{\Lambda}(p_E^2)}{\pi p_E^2}\int_{0}^{p_E}dk_E\frac{k_E^5}{k_E^2+\Sigma^2(k_E^2)}-\frac{2}{\pi}\int_{p_E}^{\infty}dk_E\frac{\alpha_s(k^2)f_{\Lambda}(k_E^2)k_E}{k_E^2+\Sigma^2(k_E^2)}.\label{eq:sd_d}
\end{align}
% where the angle integration has been done (see Ref.~\cite{Aoki1990}) as
% \begin{align}
%  \int d\Omega\frac{4p_E^2k_E^2-4(p_E\cdot k_E)^2}{(p_E-k_E)^4}=\frac{3p_E^2k_E^2-3\min(k_E^4,p_E^4)}{|k_E^2-p_E^2|\max(k_E^2,p_E^2)}.
% \end{align}
%%%with the bare diquark mass $m_{\phi}$.
%Here, we take $\Lambda_{\rm UV}=5$GeV for the ultraviolet cutoff.

% and $p_E$ denotes the momentum in the Euclidean space.
%The diagrammatic expression of the SD equation is shown in Fig.~\ref{fig:sd_d}. 
% The second term corresponds to the 4-point interaction term and the third term is the 3-point interaction term in Fig.~\ref{fig:sd_d}.
% Here, we do not consider the wave functional renormalization, since the diquark is confined in a hadron and do not produce any physical quantities. The physical quantities will be discussed in the color singlet state such as tetra-quark in our future work.

\section{Numerical Result}
\subsection{The Parameters in the Diquark Theory}
The bare mass $m_{\phi}$ and the size $R$
(inverse of cutoff $\Lambda$) are free parameters of the diquark theory. 
%We have no physical value to fix these parameters, but we can restrict them from the physical analysis.
 Since the diquark is originally made of two quarks, 
the bare diquark mass can be simply considered as twice of the quark mass. 
 In this paper, we consider two cases of the bare diquark mass. 
One is twice of the constituent quark mass, i.e., $m_{\phi}=600$ MeV. 
The other is twice of the running quark self-energy, i.e., 
$m_{\phi}(p_E^2)=2\Sigma_q(p_E^2)$, where $\Sigma_q(p_E^2)$ is determined 
by the SD equation for single quark 
\cite{Higashijima1984,Aoki1990,Yamanaka2013}.
%\begin{align}
% \Sigma_q(p^2_E)%=&m_q+i\int\f{k}{4}g^2((p-k)^2)\gamma^{\mu}T_a\frac{1}{\Slash{k}-\Sigma_q(k^2)}\gamma^{\nu}T^{b}D_{\mu\nu}^{ab}((p-k)^2)\nn
%%=&m_q+\frac{3C_{2}(\bm{3})}{4\pi^3}\int d^4k_E \frac{\alpha_s((p_E-k_E)^2)}{(p_E-k_E)^2}\frac{\Sigma_q(k_E^2)}{k_E^2+\Sigma_q^2(k_E^2)}\nn
%=&m_q+\frac{2\alpha_s(p^2)}{\pi p_E^2}\int_{0}^{p_E}dk_E\frac{k_E^3\Sigma_q(k_E^2)}{k_E^2+\Sigma_q^2(p_e^2)}+\frac{2}{\pi}\int_{p_E}^{\Lambda_{UV}}dk_E\frac{k_E\alpha_s(k_E^2)\Sigma_q(k_E^2)}{k_E^2+\Sigma_q^2(k_E^2)}.\label{eq:sd_q}
%\end{align}
%This means that the diquark is constructed by the two dressing quarks. The constant bare mass case is based on the constituent quark model like picture and the running bare mass case is the SD formalism with omitting the effect of the gluonic attraction force between two quarks. 
%The diquark should be dressed by gluon furthermore because the diquark has non-zero color charge.
%The cutoff $\Lambda$ corresponds to the diquark size in a hadron. %In this point of view, the diquark should be smaller than the hadron.
 We also consider two cases of the diquark size $R$ corresponding to 
the cutoff $\Lambda$. 
One is the typical size of a baryon $R=1$ fm, i.e., $\Lambda=200$ MeV, 
which gives the upper limit of the size (the lower limit of the cutoff).
%The diquark covers the baryon in this case. 
 The second is the typical size of a constituent quark $R\simeq0.3$ fm, 
i.e., $\Lambda=600$ MeV, 
which gives the lower limit of the size (the upper limit of the cutoff).

\subsection{The Constant Bare Mass Case}\label{constant}
\begin{figure}[htbp]
\centering
  \subfigure[$R=1$ fm ~~ (i.e., $\Lambda=200$ MeV)]
{\includegraphics[width=0.45\textwidth]{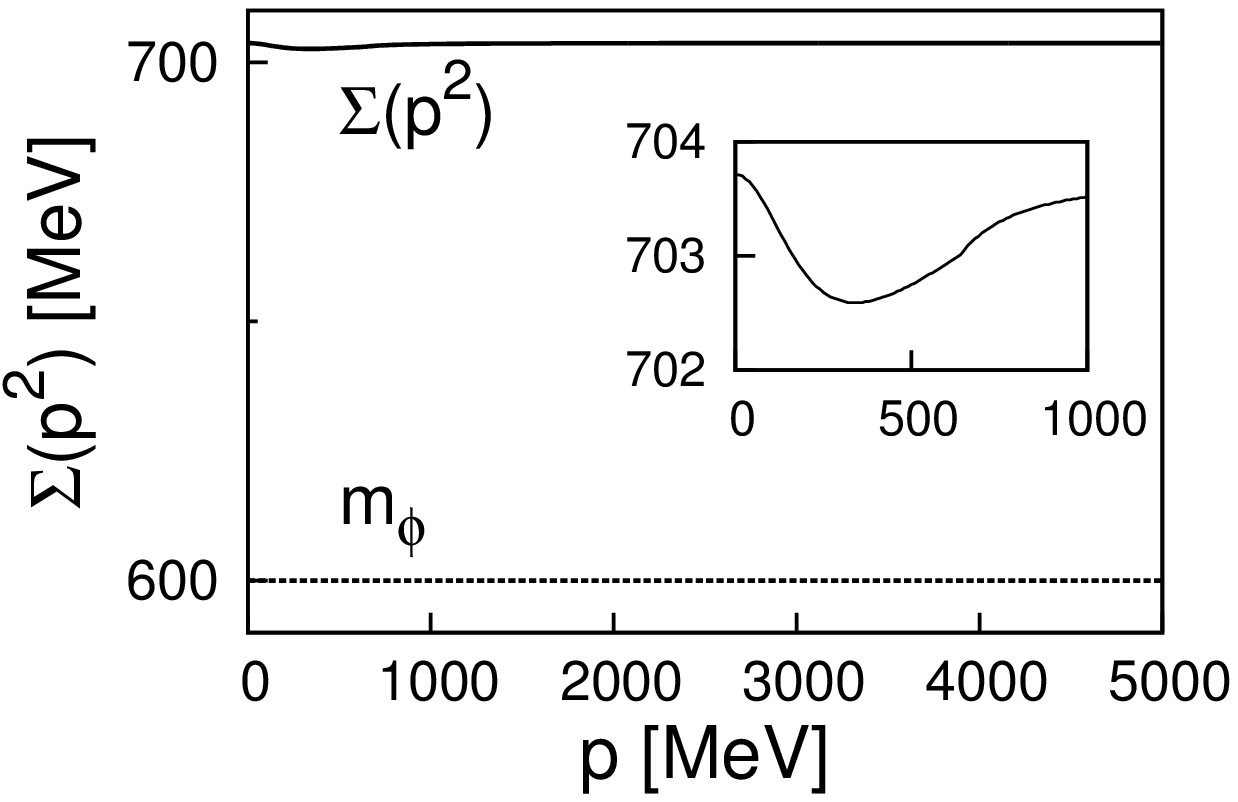}\label{fig:c2}}
  \hfill
  \subfigure[$R\simeq 0.3$ fm ~~ (i.e., $\Lambda=600$ MeV)]
{\includegraphics[width=0.45\textwidth]{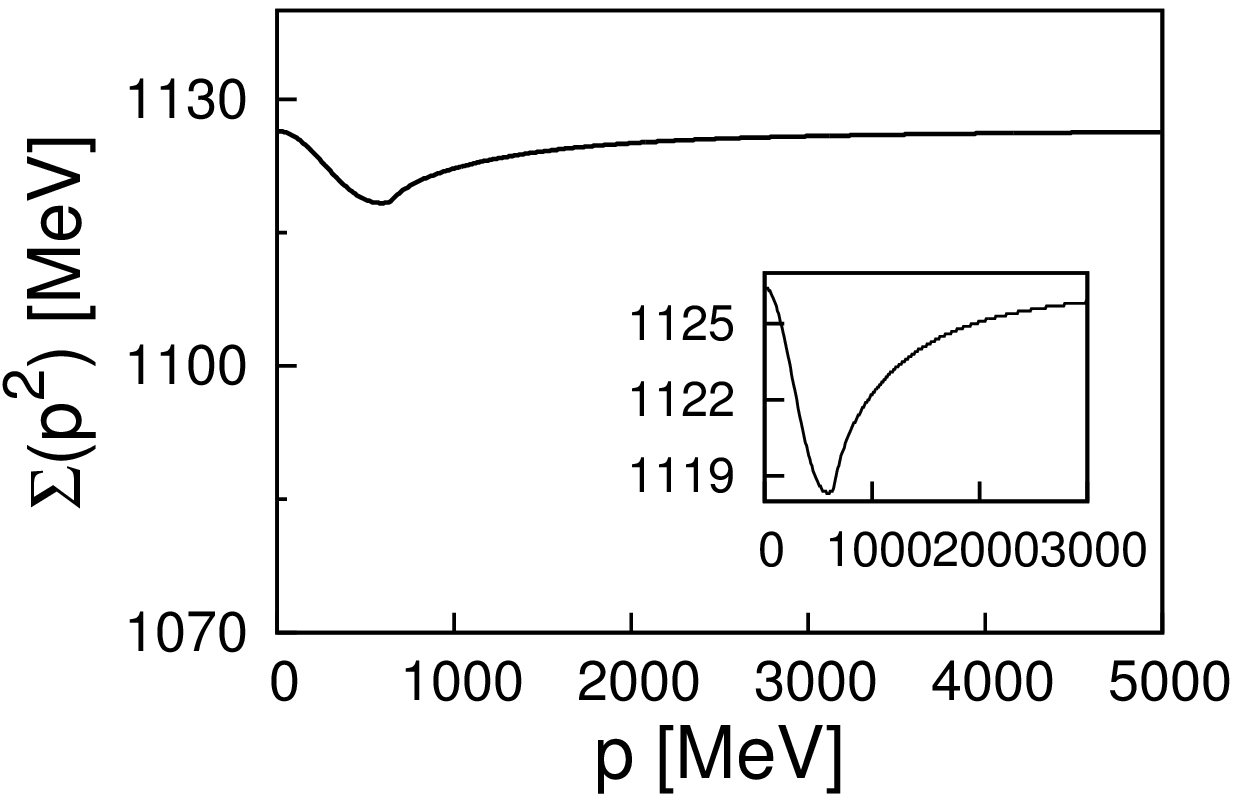}\label{fig:c6}}
\caption{The scalar-diquark self-energy $\Sigma(p^2)$ as a function of the momentum in the constant bare-mass case of $m_\phi=600$ MeV with \protect\subref{fig:c2} $R=1$ fm and \protect\subref{fig:c6} $R\simeq 0.3$ fm. In both cases, there appears a small bump structure, which is displayed in the small window. In the left figure, the original bare mass $m_{\phi}$ is plotted for comparison.}\label{fig:cm}
\end{figure}

We first show in Fig.~\ref{fig:cm} 
the case of the constant bare mass $m_{\phi}=600$ MeV 
with dependence on the diquark size $R$. %The self-energy $\Sigma(p^2)$ strongly depends on the cutoff $\Lambda$, but it 
The diquark self-energy $\Sigma(p_E^2)$ is always larger than the bare mass 
$m_{\phi}$ and almost constant except for a small bump structure 
at an infrared region. 
The value of the self-energy strongly depends on the size $R$, 
e.g., the ``compact diquark'' with $R\simeq 0.3$ fm has a large mass.

\subsection{The Running Bare Mass Case}

\begin{figure}[htbp]
\centering
  \subfigure[$R=1$ fm ~~ (i.e., $\Lambda=200$ MeV)]
{\includegraphics[width=0.45\textwidth]{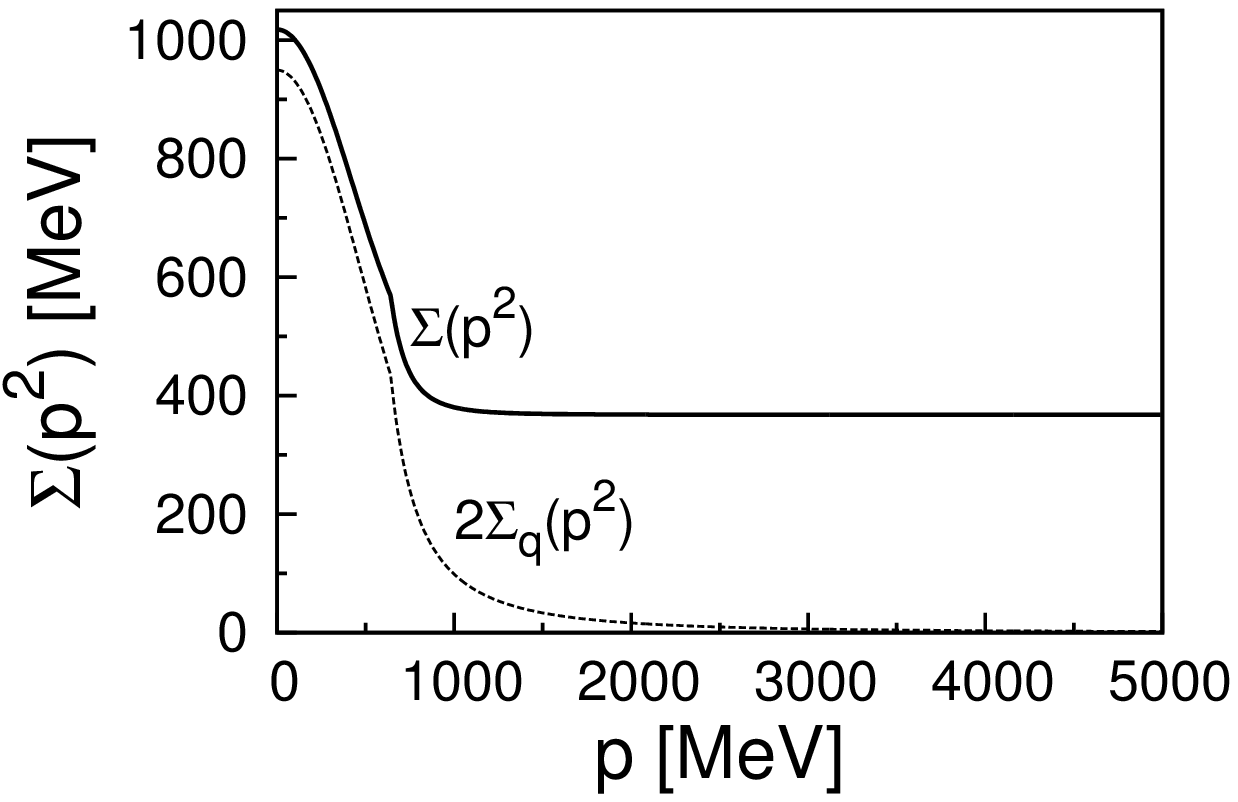}\label{fig:q2}}
  \hfill
  \subfigure[$R\simeq 0.3$ fm ~~ (i.e., $\Lambda=600$ MeV)]
{\includegraphics[width=0.45\textwidth]{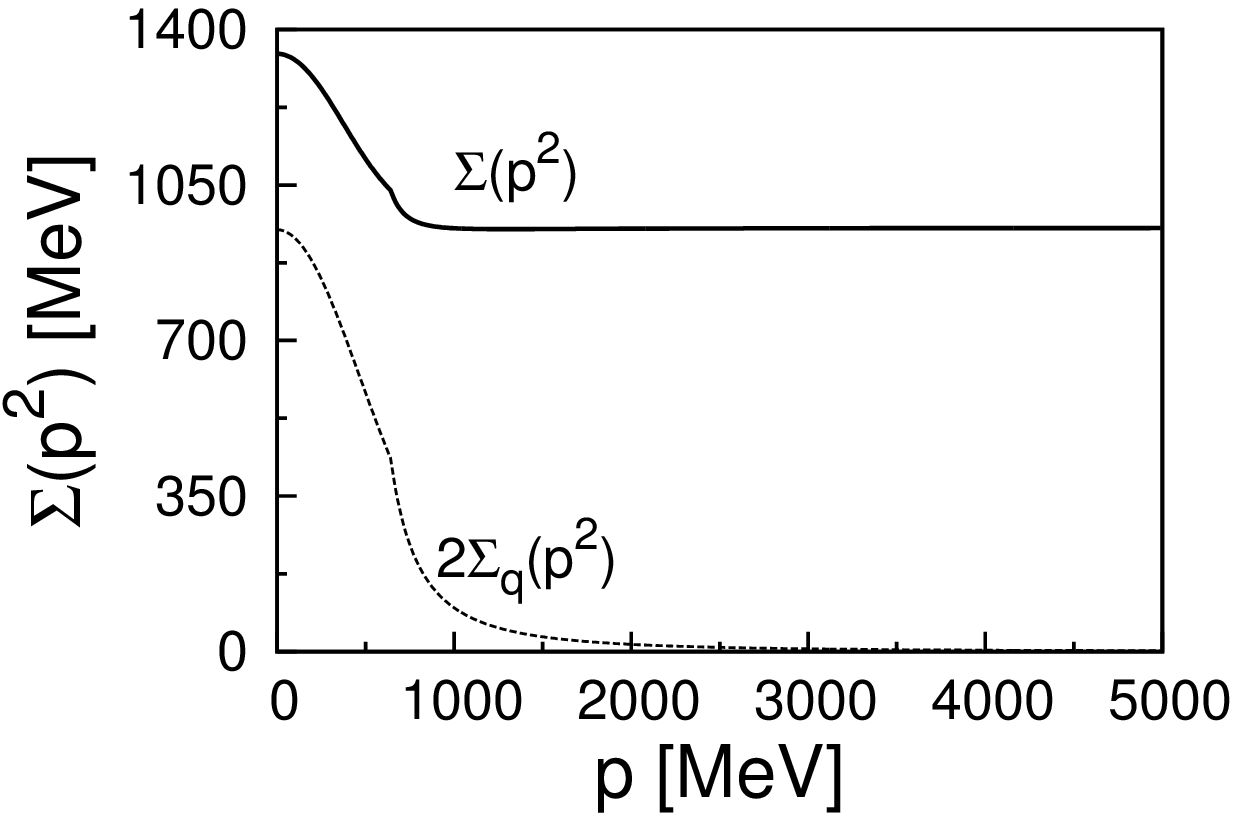}\label{fig:q6}}
  \caption{The scalar-diquark self-energy $\Sigma(p^2)$ as a function of the momentum in the running bare-mass case with \protect\subref{fig:q2} $R=1$ fm and \protect\subref{fig:q6} $R\simeq 0.3$ fm. The diquark bare mass $m_{\phi}(p^2)=2\Sigma_q(p^2)$ is also plotted with the dotted line for comparison.}\label{fig:rm}
\end{figure}

We show in Fig.~\ref{fig:rm} the case of the running bare mass 
$m_{\phi}(p_E^2)=2\Sigma_q(p_E^2)$ with dependence on $R$. 
The diquark self-energy $\Sigma(p_E^2)$ also strongly depends on 
the diquark size $R$. 
In the low-momentum region, the behavior of $\Sigma(p_E^2)$ reflects 
the running property of the bare mass,
%{\itshape  According to the discussion of the previous section~\ref{constant} (see Fig.~\ref{omit}), the 3-point interaction term would make a bump structure, but this behavior is hidden by the running bare mass, and the 4-point interaction term rise the value.}
especially in the $R=1$ fm case, the gluonic effect seems 
to be small, because of $\Sigma(p_E^2) \approx 2\Sigma_q(p_E^2)$.
In the high-momentum region, 
the diquark self-energy keeps a large finite value, 
while the bare mass $m_\phi(p_E^2)$ goes to zero. 
This suggests the mass generation of the scalar diquark 
by the gluonic radiative correction \cite{Iida2007}. 

\section{Dynamical Mass Generation without Chiral Symmetry}

\begin{figure}[htbp]
\centering
  \subfigure[$R=1$ fm ~~ (i.e., $\Lambda=200$ MeV)]
{\includegraphics[width=0.45\textwidth]{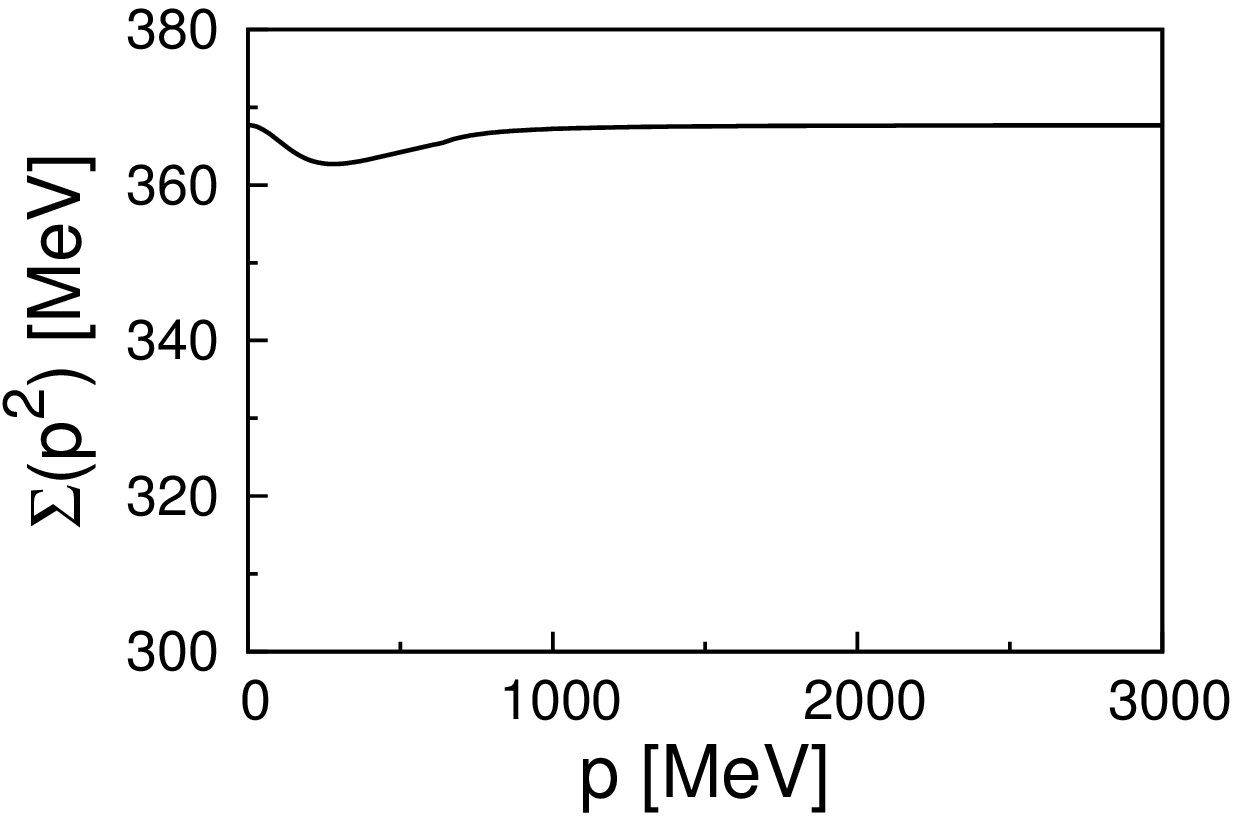}\label{fig:m2}}
  \hfill
  \subfigure[$R\simeq 0.3$ fm ~~ (i.e., $\Lambda=600$ MeV)]
{\includegraphics[width=0.45\textwidth]{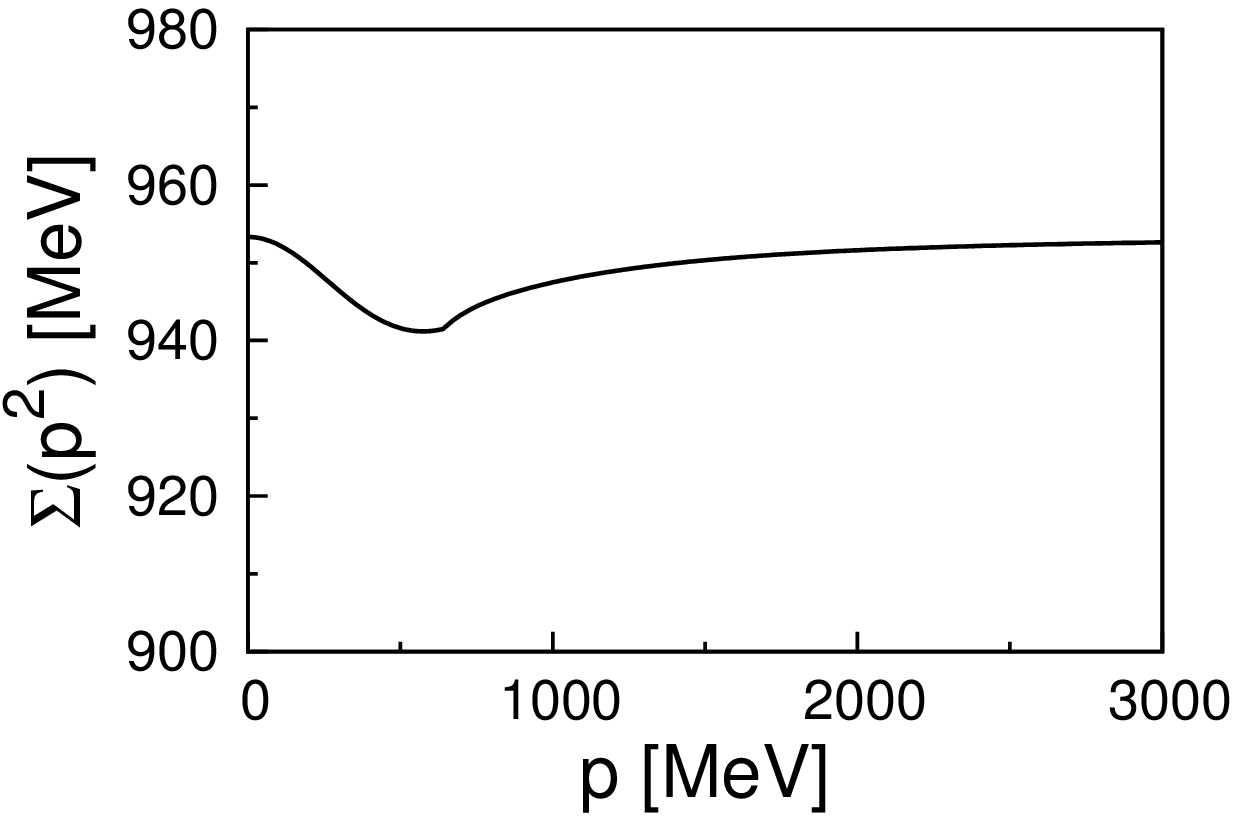}\label{fig:m6}}
\vspace*{-3mm}
  \caption{The scalar-diquark self-energy $\Sigma(p^2)$ as a function of the momentum in the massless case of $m_\phi=0$. The self-energy $\Sigma(p^2)$ is finite in both cases.}\label{fig:zero}
\end{figure}

%The mass generation in QCD without the chiral symmetry breaking.
Finally, we consider the zero bare-mass case of diquarks, $m_{\phi}\equiv 0$. 
The result is shown in Fig.~\ref{fig:zero} for the two cases of 
$R=1$ fm and $R\simeq 0.3$ fm on the diquark effective size. 
The self-energy $\Sigma(p_{E}^2)$ is always finite and takes a large value 
even for $m_{\phi} \equiv 0$. 
%As we discussed in previous section, the dynamical mass of the diquark seems to be generated.
The mass generation mechanism in QCD is usually considered in the context of 
spontaneous chiral symmetry breaking.
%It is considered that the quarks have small current mass $m_{u,d}\sim2-5$ MeV (current mass of up and down quark) in QCD Lagrangian level and the chiral symmetry is spontaneouly broken by the non-perturbative effect. The constituent quarks (up and down) have a large mass $M_{u,d}\sim300$ MeV.
% It is considered that the quarks obtain large mass by the chiral symmetry breaking. 
On the other hand, our scalar-diquark theory is composed by 
an effective scalar-diquark field $\phi(x)$ and 
does not have the chiral symmetry explicitly, 
although the original diquark is constructed by two quarks.
%We introduce the scalar diquark as a fundamental field and the Lagrangian does not have chiral symmetry. The mass of diquark should be emerged by the gluonic effect.
%Thus, we can conclude that the mass of scalar diquark is dynamically generated by the non-perturbative gluonic effect. %Since the value of self-energy strongly depends on the cutoff $\Lambda$, the

%It is known that QCD has another mechanism of mass generation. 
We consider that QCD has several dynamical mass-generation mechanism, 
even without chiral symmetry breaking.
For example, while the charm quark has no chiral symmetry, 
some difference seems to appear between current and constituent masses 
for charm quarks: the current mass is $m_c \simeq$ 1.2 GeV 
at the renormalization point $\mu=2$ GeV~\cite{Beringer2012}, 
and the constituent charm quark mass is $M_c\simeq1.6$ GeV in the quark model. 
Furthermore the gluon is more drastic case. 
While the gluon mass is zero in perturbation QCD, 
the non-perturbative effect of the self-interaction of gluons seems to 
generate a large effective mass of 0.6 GeV~\cite{Iritani2009}, 
and the lowest glueball mass is about 1.6GeV.
%The mass of gluon should be zero in the QCD Lagrangian. The mass of glueball is calculated around 1.6 GeV in lattice QCD simulation~\cite{Morningstar1999}, which can be understood as the ``constituent gluon mass'' is 0.6 GeV.
%
Although the heavy quark and gluons do not have the chiral symmetry, 
they obtain large effective mass by non-perturbative effects. 
In this sense, the scalar diquark mass can be also generated by 
the gluonic effect. 
Thus, we consider that the mass generation mechanism 
is a general property of the strong interacting theory, 
%The non-perturbative interaction can emerge dynamical mass for any particle as shown in Fig.~\ref{fig:mass}. 
and one typical mass generation is given by 
spontaneous chiral-symmetry breaking, 
which is also caused by the non-perturbative interaction.

%\begin{figure}[htbp]
% \centering
%\includegraphics[width=.5\textwidth]{mass.eps}
%\caption{The schematic picture for dynamical mass generation of the colored particle. The colored particle (solid line) interacting with the gluons (wavy line). The effective mass is emerged by the non-perturbative interaction even without the chiral symmetry.}\label{fig:mass}
%\end{figure}

\section{Summary}
%Summary will be written in here.
We have investigated the gluonic dressing effect to the scalar diquark, 
considering the size effect of the diquark in a hadron.
The non-perturbative effect is evaluated in the Schwinger-Dyson (SD) formalism 
in the Landau gauge. 
%with the renormalization group improved coupling and the Higashijima-Miransky approximation. 
%The basic technology of scalar SD formalism is improved from the quark case, such as the running coupling improving the renormalization group analysis and the Higashijima-Miransky approximation.
Since the diquark is located inside a hadron, the diquark size $R$ 
must be smaller than the hadron ($\sim 1$ fm) and 
larger than the constituent quark ($\sim 0.3$ fm).
We have considered two cases of the constant bare mass $m_{\phi}=600$ MeV and 
the running bare mass $m_{\phi}(p_E^2)=2\Sigma_q(p_E^2)$. 
The diquark self-energy strongly depends on the size $R=\Lambda^{-1}$ 
in both cases, especially the small diquark ($R\simeq 0.3$ fm, i.e., $\Lambda=600$ MeV) obtains a large effective mass by the gluonic dressing effect.

We find that the effective diquark mass is finite and large 
even for the zero bare-mass case, and the value strongly depends on 
the size $R$, which is an example of dynamical mass generation 
by the gluonic dressing effect, without chiral symmetry breaking.

\acknowledgments{
%S. I. thanks T.M. Doi, H. Iida and N. Yamanaka for useful discussion and comments.
  This work is in part supported by the Grant for Scientific Research [Priority Areas ``New Hadrons'' (E01:21105006), (C) No.23540306] from the Ministry of Education, Culture, Science and Technology (MEXT) of Japan.}
% This work is supported by the Grant for Scientific Research [Priority Areas ``New Hadrons'' (E01:21105006), (C) No.23540306] from the Ministry of Education, Science and Technology of Japan.}

% Insert the Acknowledgment text here.

% % can use a bibliography generated by BibTeX as a .bbl file
% % BibTeX documentation can be easily obtained at:
% % http://www.ctan.org/tex-archive/biblio/bibtex/contrib/doc/

%\bibliographystyle{h-elsevier}
%\bibliographystyle{h-physrev}
%\bibliographystyle{phaip}
%\bibliographystyle{aip}
%\bibliography{papers,sd,lattice,mynotes,confinement,qcd,text}

\end{document}